\documentstyle[12pt,aaspp4]{article}  

\begin{document}  
 
 
\title{\bf A Universal Profile of the Dark Matter Halo and 
the Two-point Correlation Function}  
  
\author{Taihei Yano\altaffilmark{1}
\altaffiltext{1}{Research Fellow of the Japan 
Society for the Promotion of Science.} and Naoteru Gouda}  
  

\affil{National Astronomical Observatory, 
Mitaka, Tokyo 181-8588, japan\\
E-mail: yano@pluto.mtk.nao.ac.jp, naoteru.gouda@nao.ac.jp}
  
\begin{abstract}  
We have investigated 
the relation between the two-point spatial correlation function and the 
density profile of the dark matter halo in the strongly non-linear regime.
It is well known that 
when the density fluctuation grows into dark matter halo 
whose density profile is  
$\rho \propto r^{-\epsilon}$($\frac{3}{2}<\epsilon<3$) on almost all mass 
scales, 
the two-point spatial correlation function obeys a power law with 
the power index $\gamma = 2\epsilon -3$ in the strongly non-linear regime. 
We find the form of the two-point spatial correlation function, 
which does not obey the power law 
when the power index $\epsilon $ is smaller than 
$\frac{3}{2}$, such as the density profile 
$\rho \propto r^{-1}$ around the center of the halo which is proposed by 
Navarro, Frenk \& White (1996,1997).
By using the BBGKY equation in the strongly non-linear regime, 
it is also found that velocity parameter 
$h \equiv - \langle v \rangle / \dot{a}x$ 
is  not a constant even in the strongly non-linear regime 
($\tilde{x} \equiv x/x_{nl} \rightarrow 0$) 
although it is a constant when $\epsilon > 3/2$ and then the 
two-point spatial correlation function can be regarded as the power law. 
The velocity parameter $h$
becomes 0 at the non-linear limit of $\tilde{x} \rightarrow 0$,
 that is, the stable clustering hypothesis cannot be satisfied 
when $\epsilon < 3/2$. 
\end{abstract}  

{\it Subject headings}:  
cosmology:theory-large scale structures-dark matter halo

\section{INTRODUCTION}

Formations of the large scale structures in the 
expanding universe is one of the most important
and interesting problems of cosmology.
It is generally believed that these structures have 
been formed owing to the gravitational instability.
 Hence it is very important to clarify evolutions of density fluctuations  
by the gravitational instability.
Here we consider density fluctuations of collisionless cold particles such as
cold dark matters, our interest being mainly
concentrated on effects of the self-gravity.  

In the hierarchical clustering picture, small fluctuations grow
and form small clusters of dark matter (dark halo) at first. 
Then these small dark halos merge into a larger halo or 
accrete surrounding matters and produce the larger dark matter halos.
Evolutions of density fluctuations in the non-linear regime are
especially important to understand the formation of the dark matter halos.
  
%

A central region of the dark halo profile was investigated by 
using N-body simulations in 1990's 
(Frenk et al. 1988, Navarro, Frenk \& White 1996, 1997 : hereafter NFW ).  
Hernquist (1990) investigated the elliptical galaxy analytically.
NFW and Hernquist claimed that dark halos are not well approximated  by 
isothermal spheres. Instead, the density profile is well approximated  
by the following form:

\begin{equation}
\rho \propto \frac{1}{(\frac{r}{r_{s}})^{\epsilon}
(1+\frac{r}{r_{s}})^{\mu -\epsilon}},
\label{01}
\end{equation}
where, $r_s$ is a characteristic scale of the dark halo.
The parameter, $\epsilon$, and $\mu$  are equal to 
$1$ and $4$, respectively, in the Hernquist's case, and
$\epsilon =1$, $\mu =3$ in the NFW's case.
In any case, the density profile of the central region is shallower 
than the isothermal sphere.

%

On the other hand, the two-point spatial correlation function is well used in 
the analysis of the density fluctuations of the dark matters. 
Of course, the two-point spatial correlation function in the non-linear regime is 
related to the dark halo profile if the mass weight of the dark halos
dominates in the universe, that is, almost all regions of high density 
correspond to dark halos.
Davis \& Peebles (1977) assumed the self-similar evolution and
the stability condition, that is, the mean relative physical velocity 
$\langle \dot{r} \rangle$ is equal to $0$
in the non-linear regime.
Then, they showed the index of the two-point spatial correlation function in the 
non-linear regime by using the index of the initial power spectrum, $n$:

\begin{equation}  
\xi \propto r^{-\gamma},~~~~~~~~\gamma =\frac{3(3+n)}{5+n}.
\label{03}
\end{equation}  

When we do not assume the stability condition, the index of the 
two-point spatial correlation function $\gamma$ is expressed by 
using the relative velocity parameter 
$h \equiv -\langle v \rangle/\dot{a}x$,

\begin{equation}  
\gamma =\frac{3h(3+n)}{2+h(3+n)}, 
\label{04}
\end{equation}  
where $n>-3$, and $h$ is a value of order $1$ ($0 \leq h \leq 1$)
(Yano \& Gouda (1997) ; hereafter YG) in the non-linear regime.
Here $\langle v \rangle$ is the mean relative peculiar velocity
( see eq.(\ref{1-1})) and $a$ is the scale factor.
Therefore $\gamma$ have the value around the following:
\begin{equation}  
0<\gamma <3.
\label{04-1}
\end{equation}  

We consider the case that the halo dominates in the regions of high density 
and then the density profile of the halo mainly contributes to the two-point 
spatial correlation function in the non-linear regime.
When the density profile of the dark halo is given by 
$\rho \propto r^{-\epsilon}$, then using the following equation,

\begin{equation}  
\xi(r) \sim \langle \rho (s) 
\rho (|\overrightarrow{r}+\overrightarrow{s}|)\rangle/\langle \rho \rangle ^2,
\label{04-1-1}
\end{equation}  
the two-point spatial correlation function in the strongly non-linear regime
($r \rightarrow 0$) is determined by the density profile of the dark halo and
written in the following way
(McClelland \& Silk 1977; Sheth \& Jain 1997, 
Padmanabhan \& Engineer 1998),

\begin{equation}  
\xi \propto r ^{-\gamma '},~~~~~~~~~~~\gamma '=2\epsilon -3.  
\end{equation}  

The two-point spatial correlation function diverges at the center of the 
dark halo when $\epsilon>3$ and it does so on 
$\tilde{r} \equiv r/r_{nl} \rightarrow \infty$ 
when $\epsilon <3/2$, where $r_{nl}$ is the non-linear scale of the 
density fluctuation.
Therefore $3/2 < \epsilon <3$ must be satisfied.
When the relation $3/2 < \epsilon <3$ is satisfied, 
$\gamma '$ satisfies $0< \gamma ' <3$.
This relation corresponds to the relation (\ref{04-1})
derived from eq. (\ref{04}). In other words, when 
the two-point spatial correlation function obeys the power law with $0<\gamma <3$,
the index $\epsilon $ satisfies 
$3/2 < \epsilon <3$.

%
%
When the density profile is 
$\rho \propto r^{-\epsilon}$ with $\epsilon <3/2$ 
around the center of the dark halo,
the index of the density profile have at least following two 
values in order to converge the integral.
\begin{equation}  
\rho \propto (\frac{r}{r_s})^{-\kappa(r)}~~~~~~:~
\kappa(r) =\left\{\begin{array}{ll}
\epsilon & (r<r_{s}~~~:~\epsilon <\frac{3}{2})\nonumber \\
\mu & (r> r_{s}~~~:~\mu>\frac{3}{2})
\end{array}\right..
\label{04-2}
\end{equation}

Here $r_s$ is a characteristics scale.
In this case, what is the form of the two-point spatial correlation function ?
The purpose of this paper is to obtain the two-point spatial correlation 
function in the strongly non-linear regime when the dark halos have the 
above density profile (\ref{04-2}) and the self-similar evolution 
of the dark halos is satisfied.
We also investigate what is the form of the velocity parameter $h$, 
that is, the mean relative peculiar velocity by using the BBGKY equation.

In $\S 2$, we will briefly review the power law solution of the 
two-point spatial correlation function described in YG.
We show in $\S 3$  the solution of the two-point spatial correlation function 
and the mean relative peculiar velocity
when there exists the dark halo whose density profile has the 
form of eq.(\ref{04-2}). 
We will devote to the conclusions and discussions in $\S 4$.
In the Appendix, we will consider the 
relation between the power index of the density profile of the 
dark halo and the initial power index $n$ if the 
merging or accreting far equilibrium dark 
halos could be regarded as a spherical halos.

\section{POWER LAW SOLUTIONS}
We will briefly review the power-law solutions of the two-point spatial 
correlation function $\xi$, and the mean peculiar velocity 
$\langle v \rangle$ described in YG.
We use the second BBGKY zeroth moment equation.

\begin{equation}  
\frac{\partial \xi}{\partial t}+\frac{1}{ax^2}\frac{\partial}{\partial x}
[x^2(1+\xi )\langle v \rangle]=0,
\label{1}
\end{equation}  
where, $a$ is the scale factor, $x$ is the comoving coordinate.
The mean relative peculiar velocity vector 
$\langle \mbox{\boldmath $v$} \rangle $
is defined as follows;
\begin{equation}  
\langle \mbox{\boldmath $v$} \rangle \equiv 
\frac{\langle \int 
(\mbox{\boldmath $v_2-v_1$})
f(1)f(2)d^3v_1d^3v_2 \rangle}
{\langle \int f(1)f(2)d^3v_1d^3v_2 \rangle}.
\end{equation}  
Here $\langle ...\rangle$ shows the ensemble mean for 
any pair with the fixed distance 
$x 
\equiv |\mbox{\boldmath $x_2-x_1$}| 
$ and
$f(i) \equiv f(\mbox{\boldmath $x_i, v_i$})$ ($i=1,2$) is 
the phase space density 
at the point of the phase space $(\mbox{\boldmath $x_i, v_i$})$. 
We notice that 
$\langle f(1) \rangle \equiv b(1)$ 
is the one-body probability distribution function,
$\langle f(1)f(2) \rangle \equiv b(1)b(2)+c(1,2)$
is the two-body probability distribution function, and
$c(1,2)$ is the irreducible two-point correlation function. 
Here we assume that the background universe is homogeneous and isotropic.
Then, $\langle \mbox{\boldmath $v$} \rangle \propto \mbox{\boldmath $x_2-x_1$}
\equiv \mbox{\boldmath $x$}$
and then the scalar velocity 
$\langle v \rangle \equiv
\langle \mbox{\boldmath $v$} \rangle \frac{\mbox{\boldmath $x$}}{x}$
is the only important quantity which determines $\xi$ through eq.(\ref{1}).
The mean relative peculiar velocity 
$\langle v \rangle$ can be rewritten by 

Therefore we can define the mean relative peculiar velocity 
$\langle v \rangle$ as follows:
\begin{equation}  
\langle v \rangle =
\frac{\langle \int 
\frac{\mbox{\boldmath $x$}}{x}\cdot(\mbox{\boldmath $v_2-v_1$})
f(1)f(2)d^3v_1d^3v_2 \rangle}
{\langle \int f(1)f(2)d^3v_1d^3v_2 \rangle}.
\label{1-1}
\end{equation}

We assume that the two-point spatial correlation function $\xi$ is
given by the following power-law form in the strongly non-linear regime:

\begin{equation}  
\xi = \xi_0 a^{\beta}x^{-\gamma}.
\label{2}
\end{equation}  
Then we obtain from the dimensional analysis of the equation (\ref{1})

\begin{equation}  
\langle v \rangle = -h\dot{a}x,~~~~~\beta = (3-\gamma)h,
\label{3}
\end{equation}  
where $h$ is a constant. We call this parameter $h$, the velocity parameter.
This parameter is a value of order $1$ ($0 \leq h \leq 1$)(YG)
when we consider the collisionless cold dark matters in the strongly 
non-linear regime.
We notice that the stability condition proposed by 
Davis \& Peebles (1977) corresponds to $h=1$
(stable clustering).
In this case, the collapsed object cannot be broken and clustered 
together to form the larger cluster.
Therefore, the relative physical velocity of two particles 
$\langle \dot{r}\rangle = \langle v \rangle + \dot{a}x $ 
is equal to $0$ ($ \langle v \rangle = -\dot{a}x$) 
in the non-linear regime.
We can consider the other extreme case about the clustering picture.
In this case, the smaller objects have clustered and
merged together and the completely virialized object is newly formed.
In this case, the separation of the particle is expanding with 
the Hubble velocity and then $h=0$(comoving clustering).

If the self-similarity solutions exist, the power index of the 
two-point spatial 
correlation function in the non-linear regime can be represented by 
the following form (Padmanabhan 1996; YG):

\begin{equation}  
\gamma =\frac{3h(3+n)}{2+h(3+n)},
\label{4}
\end{equation}  
where $n$ is the index of the initial power spectrum ($n>-3$). 
Therefore $\gamma$ have the value between $0$ and $3$.
Here we consider the case  
that the halo mass dominates and the index of the density profile 
$\rho \propto r^{-\epsilon}$ is the same for all mass scales. 
The relation $3-\epsilon >0$ and $3-2\epsilon <0$ must be satisfied 
so that the integral in eq.(\ref{04-1-1}) should converge.
Therefore, $\epsilon$ have the following values:
\begin{equation}  
\frac{3}{2}< \epsilon <3. 
\label{4.2}
\end{equation}  
In this case, the relation between the two-point spatial correlation function 
and the density profile is

\begin{eqnarray}  
\xi(r) &\propto& \int d^3s \rho (s) 
\rho (|\overrightarrow{r}+\overrightarrow{s}|) \nonumber \\
&=& c_2
(\frac{r}{r_{nl}})^{-2\epsilon +3}.
\label{4.1}
\end{eqnarray}  
[see McClelland \& Silk 1977; Sheth \& Jain 1997, 
Padmanabhan \& Engineer 1998]
where, 
\begin{equation}  
c_2 = \sum_{m=0}^{\infty}2(\frac{1}{3-\epsilon +2m}-\frac{1}{3-2\epsilon -2m})
(2-\epsilon )(1-\epsilon )\cdots (2-\epsilon -2m)\frac{1}{(2m+1)!}.
\label{4.1.1}
\end{equation}  

The relation between the both power indexes of the halo profile $\epsilon$
and the two-point spatial correlation
function $\gamma $ is given by

\begin{equation}  
\gamma =2\epsilon -3.
\label{4.3}
\end{equation}  
We can see from equations (\ref{4.2}) and (\ref{4.3}) that 
the index of the two-point spatial correlation function $\gamma$
have the values between $0$ and $3$.
This result is consistent with one derived from eq.(\ref{4}).

On the other hand, when we assume the density profile  
$\rho \propto r^{-\epsilon}$ with $\epsilon <3/2$ around the center of the 
dark halo,
what is the form of the two-point spatial correlation function?
We consider the relation between the two-point spatial correlation function 
and this density profile in the next section.

\section{UNIVERSAL DENSITY PROFILE}

We consider the form of the two-point spatial correlation function
when the density fluctuations grow into the dark halos 
which  have the shallower density profile $\rho \propto r^{-\epsilon}$
 with $\epsilon < 3/2$ around the center.
In this case, the density profile cannot have the single power law 
with $\epsilon<3/2$ in order to converge the 
integral about the two-point spatial correlation function in the 
$\tilde{r} \equiv r/r_{nl} \rightarrow \infty$ limit.

Navarro, Frenk \& White (1996,1997) 
and Hernquist (1990) claimed that 
the halo density profiles have the following form: 

\begin{equation}  
\rho \propto \frac{1}{(\frac{r}{ r_{s}})^{\epsilon}
(1+\frac{r}{r_{s}})^{\mu -\epsilon}},
\label{5}
\end{equation}  
where $\mu >3/2$ must be satisfied because the integral of the density 
should converge (eq. (\ref{5})).
The characteristic scale $r_{s}$ can be rewritten by $\lambda r_{nl}$,
where $\lambda$ is a constant.
For easy treatment, we assume that the density profile is written
in the following way:

\begin{equation}  
\rho = \lambda ^{-\mu}(\frac{r}{\lambda r_{nl}})^{-\kappa(r)}~~~~~~:~
\kappa(r) =\left\{\begin{array}{ll}
\epsilon & (r<\lambda r_{nl}~~:~\epsilon<3/2)\nonumber \\
\mu & (r>\lambda r_{nl}~~:~\mu>3/2)
\end{array}\right..
\label{5.1}
\end{equation}  
Here we consider the case of $\lambda <1$ and normalize the density 
such as $\rho =1$ at $r=r_{nl}$,
although the same results are obtained in general for $\lambda >1$. 
The two-point spatial correlation function is calculated by the following
form:

\begin{equation}  
\xi(r) \propto \int d^3s \rho (s) 
\rho (|\overrightarrow{r}+\overrightarrow{s}|). 
\label{5.2}
\end{equation}  
Here we have assumed that the halos are formed in the non-linear 
regime and their density profiles have the same form 
independently of mass scales as we show in eq.(\ref{5.1}), that is, 
we have assumed that 
their halos evolve self-similarly because many authors such as 
Jain \& Bertschinger (1996), Colombi, Bouchet, \& Hernquist (1996),
Yano \& Gouda (1998)
have investigated the 
self-similarity about the two-point spatial correlation function and then the 
self-similarity is widely believed.
In this case, each halo should have the 
self-similarity in the same way as 
the two-point spatial correlation function in the non-linear regime. 
Furthermore, the correlations with the other halos do not affect
the value of the two-point spatial correlation function in the 
non-linear regime 
because we consider the non-linear regime 
($\tilde{x}\equiv  x/x_{nl} \rightarrow 0$,
where $x_{nl}$ is the non-linear scale written in the comoving coordinate),
 and then almost all pairs of the 
two points with the distance  $\tilde{x} \ll 1$ are included in one halo.
Therefore, 
it is enough to consider 
the two-point spatial correlation function within 
only one halo in order to obtain the correlation function at 
$\tilde{x} \ll 1$.
Then, the two-point spatial correlation function in the non-linear regime 
can be derived as follows:

\begin{eqnarray}  
\xi(r) &\propto& \int d^3s \rho (s) 
\rho (|\overrightarrow{r}+\overrightarrow{s}|) \nonumber \\
&\propto& \frac{\lambda ^{-2\mu}}{r_{nl}^3} \int_{halo} ds d\cos \theta s^2 
(\frac{s}{\lambda r_{nl}})^{-\kappa(r)}
(\frac{y}{\lambda r_{nl}})^{-\kappa(y)}
\nonumber \\
&\sim& c_2
(\frac{r}{r_{nl}})^{-2\epsilon +3}
+\frac{1}{3-2\epsilon}\lambda ^{3-2\epsilon}
+\frac{1}{2\mu -3}\lambda ^{3-2\mu}
+O((\frac{r}{r_{nl}})^2)
\nonumber \\
&\equiv&
c_1+c_2 (\tilde{r})^{3-2\epsilon}
+O(\tilde{r}^2) 
\label{6}
\end{eqnarray}  
where $y \equiv \sqrt{r^2 +s^2 +2rs \cos \theta}$, $\theta$ is the 
angle between the direction of  
$\overrightarrow{r}$ and $\overrightarrow{s}$, 
$c_1$ and $c_2$ are  constants, and  
$c_1 = \lambda ^{3-2\epsilon}/(3-2\epsilon)
+\lambda ^{3-2\mu}/(2\mu -3)$ is a positive value
because we consider the case that $\epsilon < 3/2$, and $\mu >3/2$.
As a result, we can see that in the case of $\epsilon <3/2$, 
the two-point spatial correlation function cannot obey the 
power-law in the non-linear regime.
When the density is continuous at $r=r_s$ as shown in eq.(\ref{5.1}), 
the term of $O(\tilde{r})$ 
is eliminated. 
On the other hand,
the term of $O(\tilde{r}^2)$ 
remain. However, there is the possibility that the term of $O(\tilde{r}^2)$
or the higher terms are eliminated when the density profiles are 
smoothly continuous at $r=r_s$. In the following, we investigate the 
parameter $h$ when the two-point spatial correlation function can be expressed by 
eq.(\ref{6}).

 
Here we put the two-point spatial correlation function $\xi$ and the velocity 
parameter $h$ in the following way:

\begin{eqnarray}  
\xi &=& c_1 +c_2 a^{\gamma \alpha}\tilde{x}^{-\gamma} \nonumber \\
h &=& d_1 +d_2 a^{-\nu \alpha}\tilde{x}^{\nu},
\label{7}
\end{eqnarray}  
where $\tilde{x}\equiv x/x_{nl}$, and
$\alpha$ is a constant value for the scaling.
Here $c_1$, $c_2$, $d_1$, and $d_2$ are constants.
When the self-similarity of the two-point spatial correlation function 
is satisfied, $\alpha$ is equal to $2/(n+3)$ (DP, YG).
The index $\gamma$ satisfies $-\gamma=3-2\epsilon$, ($\epsilon < 3/2$)
(see eq.(\ref{6})).
We define the velocity parameter in the same way as the power-law case 
($\langle v \rangle = -h\dot{a}x$).
Then the BBGKY equation (\ref{1}) reduces to

\begin{equation}  
\frac{\dot{a}}{a}\gamma \alpha c_2 a^{\alpha \gamma} \tilde{x}^{-\gamma}
= \frac{\dot{a}}{a}\frac{1}{x^2}\frac{\partial}{\partial x}
[x^3 (c_1 +c_2 a^{\gamma \alpha }\tilde{x}^{\gamma})
(d_1 +d_2 a^{-\mu \alpha }\tilde{x}^{\mu})],
\end{equation}  
and then, this is rewritten as follows:

\begin{eqnarray}   
\gamma \alpha c_2 a^{\alpha \gamma} \tilde{x}^{-\gamma}
&=& 3 c_1 d_1 + (\nu +3) c_1 d_2 a^{-\nu \alpha}\tilde{x}^{\nu}
+(-\gamma +3)d_1 c_2 a^{\gamma \alpha}x^{-\gamma}
 \nonumber \\
&& +(-\gamma +\nu +3)c_2 d_2 a^{(\gamma -\nu) \alpha}\tilde{x}^{-\gamma +\nu}.
\label{8}
\end{eqnarray}   

Since we consider $\epsilon < \frac{3}{2}$,
the relation $-\gamma =3-2\epsilon > 0$ is satisfied.
Therefore, the left side of equation (\ref{8}) is $0$
in the non-linear limit ($ \tilde{x} = x/x_{nl} \rightarrow 0$).
As a result, $d_1$ must be $0$ because $c_1$ is a positive value.
Then the first and third terms of the right side of the equation (\ref{8}) 
vanish. Therefore in the limit of $\tilde{x} \rightarrow 0$,
the second term 
of the right side is the same order of the left side. 
So, $\nu$ is equal to $-\gamma$.
Then, the last term of the right side is higher order.
Thus we obtain the following relation 

\begin{equation}  
\gamma \alpha c_2 = (\nu +3) c_1 d_2, 
\label{9}
\end{equation}  
or equivalently,
\begin{equation}  
\nu = \frac{-3 c_1 d_2}{c_1 d_2 + \alpha c_2}.
\label{10}
\end{equation}  

Here  $\alpha$ and $c_2$ are order of 1. $\gamma$ and $\nu+3$ are the 
same order.  $c_1$ is the value of the two-point spatial correlation function in the 
non-linear limit ($\xi (0)$). Then $c_1$ is greater than 1.
If we assume $c_1 \sim 10^2$, then $d_2 \sim 10^{-2}$.

Finally, we find that 
the velocity parameter is written in the following form:

\begin{equation}  
h=d_2 a^{-\nu \alpha}\tilde{x}^{\nu}~~~~~~:~
\nu =3-2\epsilon~~~  ( \epsilon<\frac{3}{2} ),~~~~~ d_2 \ll1.
\label{10.1}
\end{equation}  
The velocity parameter $h$ is zero in the 
non-linear limit. This means that the stability condition ($h=1$) cannot be 
satisfied at least in the case of $\epsilon <3/2$.

We obtain the form of the velocity parameter as shown in eq. (\ref{10.1}), and 
this velocity parameter is related with the two-point spatial correlation function 
through $\nu$. However, we can not obtain the values of 
$d_2$ and $\nu$ from the second BBGKY
zeroth moment equation. If we would like to do so, we must solve 
higher moment equations. 

Here we consider the meaning of the velocity parameter $h$ again.
The parameter $h$ is defined as follows:

\begin{equation}  
\langle v \rangle = -h \dot{a}x
=\frac{\langle \int 
\frac{\mbox{\boldmath $x$}}{x}\cdot(\mbox{\boldmath $v_2-v_1$})
f(1)f(2)d^3 v_1d^3 v_2 \rangle }
{\langle \int f(1)f(2)d^3 v_1d^3 v_2 \rangle}. 
\end{equation}  
This velocity parameter is derived from an ensemble mean of the 
relative peculiar velocity, and can be regarded as a spatial mean of  
all regions of the universe when the distance of the two-points is much 
smaller than the horizon scale 
and we can get many samples of the pairs. 
Then if we average the relative peculiar velocity within a part of the
region of the universe,
this value is generally different from the mean value 
averaged within all regions of the universe.
Here we define the following parameter $h_{local}$;

\begin{equation}  
\langle v \rangle _{local} \equiv -h_{local}\dot{a}x
\equiv \frac{\langle \int 
\frac{\mbox{\boldmath $x$}}{x}\cdot(\mbox{\boldmath $v_2-v_1$})
f(1)f(2)d^3 v_1d^3 v_2 \rangle _{local}}
{\langle \int f(1)f(2)d^3 v_1d^3 v_2 \rangle _{local}},
\label{vlocal}
\end{equation}  
where the subscript ``local'' means that 
the average is taken only in a dark halo
for any pair with the fixed distance $x= |\mbox{\boldmath $x_2-x_1$}|$. 
When a halo becomes isolated and virialized,
this halo might have the value of the parameter $h_{local}=1$
even when the stability condition cannot be satisfied ($h \ne 1$)
(see Appendix).
It should be noted that 
the parameter $h_{local}=1$ does not necessarily mean $h=1$.

We obtain the result of $h=0$ in the non-linear limit as we can see from 
eq. (\ref{10.1}).
Then if $h_{local}>0$ 
in a dark halo, 
there must exist another halo in which $h_{local}<0$.
Because if all halos have $h_{local}>0$, $h$ should be larger than $0$ 
and it is inconsistent with the condition of $h=0$.


Here we assume that the density profile 
is determined independently of the initial conditions.
In this case, a kind of ``relaxation'' process must work for the 
density profile.
Then, the velocity field of each halo also might relax to a 
common state independently of the initial condition.
Therefore, all halos might have the same $h_{local}$.
Since we obtain $h=0$ in the nonlinear limit, 
$h_{local}$ for all halos must be $0$.

Only 
if the power index $\epsilon$ of the density profile is equal to $0$,
the both conditions of $h=0$ and $h_{local}=0$ for all halos can 
be satisfied 
(see eq.(\ref{4-8-9}) in Appendix).
If $\epsilon \ne 0$, as appeared in the NFW density profile, then 
$h_{local} \ne 0$. So it is impossible that all halos with 
$\epsilon \ne 0$ have the same value of $h_{local} \ne 0$ independently of 
initial conditions.

\section{CONCLUSIONS AND DISCUSSION}  
We have investigated 
the relation between the two-point spatial correlation function and the 
density profile of the dark matter halos 
in the strongly non-linear regime.
It is found that when the density fluctuations evolve to the dark matter halos 
whose density profile is  $\rho \propto r^{-\epsilon}$ ($\epsilon<3/2$) 
around the center of the halo, 
the two-point spatial correlation function cannot obey the power law.
Furthermore we find by using the BBGKY equation that the velocity 
parameter $h \equiv - \langle v \rangle / \dot{a}x$ 
is not a constant in the case of the shallower density profile 
($\epsilon <3/2$) although $h$ is a constant in the case of 
$\epsilon>3/2$.
The velocity parameter 
$h$ becomes 0 at the non-linear limit,
 that is, the stable clustering hypothesis cannot be satisfied at least 
in the case of $\epsilon<3/2$. 




We consider the power spectrum when the index of the 
density profile is shallower than $3/2$:

\begin{eqnarray}  
P(k) &=& \int \xi (r) e^{ikx}d^3r \nonumber \\
&=&\int [c_1 +c_2(\frac{r}{r_{nl}})^{\nu}] \frac{\sin kr}{kr}r^2dr \nonumber \\
&\sim& \frac{c_1}{3}k^{-3}+\frac{c_2 r_{nl}^{-\nu}}{3+\nu}k^{-(3+\nu)}.
\label{11}
\end{eqnarray}  
The index $\nu = 3-2\epsilon$ of the power spectrum is 

\begin{equation}  
\frac{\partial \ln P}{\partial \ln k}
=-3 -\frac{\nu K k^{-\nu}}{1+ K k^{-\nu}}~~(>-3),
\label{12}
\end{equation}  
where $K \equiv 3c_2 r_{nl}^{\nu}/(3+\nu)c_1(<0)$. 
In the case of the power index $\frac{\partial \ln P}{\partial \ln k}<-3$,
the power spectrum evolves and the power index changes to the value $-3$
which is decided by the catastrophe theory (Gouda 1998).
However, as we see from eq.(\ref{12}), the index of the power law is 
larger than $-3$ when the power spectrum is fitted by the 
single power law in large $k$ although the power spectrum does not obey the 
power law in real.


In the paper of the NFW, it is proposed that the density profile is 
independent of the initial condition and the index of the density 
profile $\epsilon$ is equal to $1$. 
Here we assume that the NFW density profile is realized in almost all halos 
on all mass scales of the halo.
In this case, a kind of ``relaxation'' process must work for the density 
profile.
Then the velocity field also might be relaxed in all halos.
Therefore all halos might have the same $h_{local}$.
Since we obtain $h=0$ in the non-linear limit, the parameter $h_{local}$
must be $0$ if $h_{local}$ is common for all halos.
As we can see from the argument in Appendix ( eq.(\ref{4-8-9})),
$h_{local}$ becomes $0$ only when the index of the 
density profile $\epsilon =0$.
Primack \& Bullock (1998) proposed by simulations that the density profile in
the center of the dark halos have the index with $\epsilon \approx 0.2$.
This value is different from the value proposed by NFW.
Furthermore this value $\epsilon \approx 0.2$ is nearly equal to $0$.
Therefore there exists the possibility that all halos have the 
same $h_{local}$ and its value is $0$.
If the index of the density profile of the dark halos have the value 
of $\epsilon =1$ that is proposed by NFW, $h_{local}$ must not be $0$ 
in the central region of the dark halo.
In order to become $h=0$, 
there might exist another halo in which $h_{local}<0$,
that is, each halo might have a different velocity field.
Here it is hard to believe that each halo has independently the same index
with $\epsilon =1$ although each halo has the different velocity field.
Then, it is natural that the index $\epsilon$ also depends on the initial 
condition
or $\epsilon = 0$ for all halos on all mass scales.

On the other hand, Huss, Jain, \& Steinmetz (1998) have claimed that the 
density profiles of the isolated object have a nearly universal characteristic 
shape independently of the initial condition or the formation history.
This means that 
the halo which is saved from the merging or the accretion for enough time
to become equilibrium might have the index $\epsilon =1$. 
If the dark halos with the index $\epsilon =1$ 
whose density profile is determined independently of the initial condition 
dominate in the universe, 
each halo might have the same $h_{local}$ and $h_{local}$ is larger than $0$
(see eq.(\ref{4-8-9}) in Appendix), and then $h$ cannot be equal to $0$.
Therefore, such halos can not dominate because $h$ must be $0$ 
in the non-linear limit.
In order to become $h=0$, we can consider that such virialized and equilibrium
halos do not dominate and far equilibrium and distorted halos dominate 
in the universe.
And then, the far equilibrium halos, that is, the merging or accreting
distorted halos mainly determine the self-similarity of the 
two-point spatial correlation function. 

We consider the case that 
almost all halos are merging or accreting and are not equilibrium.
Such halos can not be spherical in a real situation.
However, we assume the distorted dark halos could be regarded 
as spherical halos.
Then we can regard $\rho$ as a function of $r$ only.
In this case we investigate the density profile of the dark halos,
the two-point spatial correlation function,
and the velocity parameter as a function of the index of the 
initial power spectrum as we show in the Appendix.
We assume that the self-similarity of the  
dark halos satisfies and that $h_{local}$ is equal to $1$.
In this assumption, the index of the density profile becomes 
$\epsilon = 3(n+3)/(n+5)$ from eq.(\ref{4-9}) in the Appendix. 
In the case of $n>-1$, that is, the index of the density profile 
is $3/2 < \epsilon <3$ because $\epsilon = 3(n+3)/(n+5)$.
Furthermore the velocity parameter is $h=(n+1)/2(n+3)$ and 
the index of the two-point spatial correlation 
function is $\gamma=3h(n+3)/[2+h(n+3)]=3(n+1)/(n+5)$.
In the case of $n<-1$, the index of the density profile is
$\epsilon <3/2$. In this case, the power index of the two-point spatial correlation
function in the non-linear limit is $0$ 
($\xi = c_1 + c_2 (r/r_s)^{3-2\epsilon}$),
and the velocity parameter in the non-linear limit is $0$
($h=d_2 \tilde{x}^{3-2\epsilon}$). 

\section{APPENDIX}
The merging  or accreting dark halos are distorted and far equilibrium.  
In this Appendix, 
we show the relation between the power index of the density profile of 
the halo and the initial power index $n$ 
if such distorted dark halos could be regarded as a spherical halos 
and we can obtain the 
density profile which depends only on the distance $r$ from the center of 
the halo. 
When we assume that the self-similar evolution and the hierarchical clustering 
are satisfied, 
the density of the dark halo can be written as follows:

\begin{equation}  
\rho (r)= \delta _M (\frac{r}{r_s})^{-\epsilon},
\label{4-1}
\end{equation}  
where $\delta_M \propto a^{-3}$ is the normalization factor and 
$r_s \propto a^{2/(n+3)}a =a^{(n+5)/(n+3)}$ 
is the characteristic size of the dark halo
(Peebles (1980), Syer \& White (1997)). 
The continuity equation of the matter in the halos is given by 

\begin{equation}  
\frac{\partial \rho }{\partial t} + \nabla \cdot
 ( \rho \bar{\mbox{\boldmath $V$}})=0,
\label{4-1-0}
\end{equation}  
where  $\bar{\mbox{\boldmath $V$}}$ is the physical 
streaming velocity. 

\begin{equation}  
\bar{\mbox{\boldmath $V$}}
=\frac{\int \mbox{\boldmath $V$} f(2)d^3v_2}{\int f(2)d^3v_2},
\end{equation}  
where  $\mbox{\boldmath $V$}$ is the physical velocity and 
$f(2)$ is the phase space density.
Here we express the streaming velocity by using a newly defined parameter
$h_{stream}$

\begin{equation}  
\bar {V}_r \equiv \bar{\mbox{\boldmath $V$}}
\cdot \hat{\mbox{\boldmath $r$}}
=\bar{v}(x)+\dot{a}x  
\equiv (1-h_{stream})\dot{a}x,
\label{4-7-1}
\end{equation}  
where $x$ is the separation from the center of the dark halo
and  $\bar{v}(x)$ is the peculiar streaming velocity. 
Hence the parameter $h_{stream}$ is expressed by the following form,

\begin{equation}  
-\dot{a}h_{stream}x=\frac{\langle \int 
\frac{\mbox{\boldmath $x$}}{x}\cdot\mbox{\boldmath $v_2$}
f(2)d^3v_2 \rangle_x}
{\langle \int f(2)d^3v_2 \rangle_x},
\label{4-7-2}
\end{equation}  
where $\langle ...\rangle_x$ means the average in a dark halo 
for the points which have 
the fixed distance $x$ from the center of the dark halo.
If the radial streaming velocity around the center of 
the dark halo is vanished due to the virialization, 
the following relation is satisfied;
($\bar {V}_r \equiv \bar{\mbox{\boldmath $V$}}
\cdot \hat{\mbox{\boldmath $r$}}=0$). Then in this case 
$h_{stream}$ is equal to $1$
as we see from eq.(\ref{4-7-1}). 

The peculiar streaming velocity $\bar{v}$ corresponds to 
the mean peculiar velocity of the two points, one of which is fixed to 
the center of the dark halo.
Therefore $\bar{v}$ is different from $\langle v \rangle _{local}$,
which is defined in eq.(\ref{vlocal}), in general.
However,
there is the possibility that 
$\bar{v}$ is equal to $\langle v \rangle _{local}$, that is,
$h_{stream}$ is equal to $h_{local}$.
For example, 
we consider the case of 
$\bar{\mbox{\boldmath $v_i$}} \propto \mbox{\boldmath $x_i$}$
where $\mbox{\boldmath $x_i$}$ is the radial vector of the i-th point
from the center of the halo.
This relation might be satisfied when the 
halo is isolated and virialized.
In this case we can show $h_{stream}=h_{local}$ as follows:
\begin{eqnarray}  
-\dot{a}h_{local}x &=&
\frac{\langle \int 
\frac{\mbox{\boldmath $x$}}{x}\cdot(\mbox{\boldmath $v_2-v_1$})
f(1)f(2)d^3v_1 d^3v_2 \rangle_{local}}
{\langle \int f(1)f(2)d^3v_1 d^3v_2 \rangle_{local}}\nonumber \\
&=&
\langle 
\frac{\mbox{\boldmath $x$}}{x}\cdot(
\bar{\mbox{\boldmath $v_2$}}-\bar{\mbox{\boldmath $v_1$}})
\rangle_{local}\nonumber \\
&=&
\langle 
\frac{\mbox{\boldmath $x$}}{x}\cdot(
\bar{\mbox{\boldmath $v_2$}}-\bar{\mbox{\boldmath $v_1$}})
\rangle_{x}\nonumber \\
&=&
\langle 
\frac{\mbox{\boldmath $x$}}{x}\cdot
\bar{\mbox{\boldmath $v_2$}}
\rangle_{x}
-
\langle 
\frac{\mbox{\boldmath $x$}}{x}\cdot
\bar{\mbox{\boldmath $v_1$}}
\rangle_{x}
\nonumber \\
&=&
\langle 
\frac{\mbox{\boldmath $x$}}{x}\cdot
\bar{\mbox{\boldmath $v_2$}}
\rangle_{x}
\nonumber \\
&=&
\frac{\langle \int 
\frac{\mbox{\boldmath $x$}}{x}\cdot\mbox{\boldmath $v_2$}
f(2)d^3v_2 \rangle_x}
{\langle \int f(2)d^3v_2 \rangle_x}\nonumber \\
&=&
-\dot{a}h_{stream}x
\label{4-7}
\end{eqnarray}  
In the above equations we use the following arguments;
Since $\bar{\mbox{\boldmath $v_i$}} \propto \mbox{\boldmath $x_i$}$,
the relation $
\bar{\mbox{\boldmath $v_2$}} -\bar{\mbox{\boldmath $v_1$}} 
\propto \mbox{\boldmath $x_2 -x_1$} = \mbox{\boldmath $x$}$
is also satisfied for any two-points in a halo.
Therefore,
if we consider the average for the fixed $\mbox{\boldmath $x_1$}$, then
 the mean value 
of the relative peculiar velocity 
for the fixed $\mbox{\boldmath $x_1$}$ in a halo
is the same 
irrespectively of the place of $\mbox{\boldmath $x_1$}$.
%
%
Since we consider the strongly non-linear limit
($\tilde{x} \rightarrow 0$),
we can neglect the effect of the boundary of the halo,
which would be appeared when $\mbox{\boldmath $x_1$}$ 
is placed near the boundary.
Then, we can change $\langle ...\rangle _{local}$ into 
the average for any fixed $\mbox{\boldmath $x_1$}$.
Here we put the $\mbox{\boldmath $x_1$}$ on the center of the halo.
Hence $\langle ...\rangle _{local}$ is exchanged into 
$\langle ...\rangle _{x}$ in which  $\mbox{\boldmath $x_1$}$
is fixed to the center of the halo 
as we defined in eq.(\ref{4-7-2}).
Furthermore we find easily that 
$\langle 
\frac{\mbox{\boldmath $x$}}{x}\cdot
\bar{\mbox{\boldmath $v_1$}}
\rangle_{x}=0$.

From eq.(\ref{4-1}), the density $\rho$ is written by the following form,
\begin{equation}  
\rho \propto a^{\epsilon(5+n)/(n+3)-3}r^{-\epsilon}.
\label{4-8-8}
\end{equation}  

We put eq.(\ref{4-8-8}) into eq.(\ref{4-1-0}), and then we obtain
\begin{equation}  
\epsilon = \frac{3h_{stream}(n+3)}{2+h_{stream}(n+3)}.
\label{4-8-9}
\end{equation}  

It is found that 
when $h_{stream}=0$ ($h_{local}=0$),
the index of the density $\epsilon$ must be $0$.
Furthermore, when $h_{stream}=1$, the value of the
index corresponds to that shown in Syer \& White(1997). 
Syer \& White(1997) shows the model that 
the density profile is generated by tidal stripping of small halos as 
they merge with larger objects.

In our model, we can estimate $h$ or 
the power index of the two-point spatial correlation function
as a function of initial condition $n$.

In the case of $3/2 < \epsilon <3$, as we can see in $\S 3$, 
we can consider that the two-point spatial correlation function obeys 
the power-law.
Furthermore, when the self-similarity is satisfied, the power index of the 
two-point spatial correlation function can be expressed by eq.(\ref{4}).
On the other hand, when we assume that $h_{local}$ is equal to $1$,
the index of the density profile $\epsilon$ is expressed 
from eq.(\ref{4-8-9}) by the following form; 

\begin{equation}  
\epsilon = \frac{3(n+3)}{n+5}.
\label{4-9}
\end{equation}  
Here $n$ must be greater than $-1$ because $3/2 < \epsilon < 3$.
On the other hand, from eqs.(\ref{4}) and (\ref{4.3}),
$\epsilon$ is given by
 
\begin{equation}  
\epsilon = \frac{\gamma +3}{2}=\frac{3h(n+3)+3}{2+h(n+3)}
\label{4-10}
\end{equation}  
From eqs. (\ref{4-9}) and (\ref{4-10}), 
we obtain the form of the velocity parameter $h$
as a function of the index of the initial power spectrum $n$ for $n > -1$.

\begin{equation}  
h=\frac{n+1}{2(n+3)}.
\label{4-11}
\end{equation}  

We put  eq.(\ref{4-11}) into eq.(\ref{4}), and then
we obtain 

\begin{equation}  
\gamma = \frac{3(n+1)}{n+5}.
\label{4-12}
\end{equation}  

In the case of the $0 < \epsilon < 3/2$, the index of the two-point spatial 
correlation function in the non-linear limit ($\tilde{x} \rightarrow 0$)
 is equal to 
$0$ because $\xi=c_1 + c_2 (\tilde{x})^{3-2\epsilon}$ with $3-2\epsilon >0$.
Furthermore, the velocity parameter $h$ is equal to $0$ in the nonlinear limit
($h=d_2 \tilde{x}^{3-2\epsilon}$). 
On the other hand, when $h_{stream}=h_{local}=1$ is assumed, the index 
of the density profile is $\epsilon =3(n+3)/(n+5)$.

\acknowledgments  
We would like to thank Y. Fujita and M. Nagashima for useful discussion.
We are also grateful to F. Takahara and M. Sasaki for useful suggestions.
This work was supported in part by 
Research Fellowships of the Japan Society for the Promotion of Science 
for Young Scientists (No.4746) and in part by the grant-in-Aid for
Scientific Research (No.10640229) from the Ministry of Education, Science,
Sports and Culture of Japan.

\end{document}